\documentclass[prl,showpacs,preprintnumbers,twocolumn,amsmath,amssymb,superscriptaddress]{revtex4-1}

\usepackage[T1]{fontenc}
\usepackage[applemac]{inputenc}
\usepackage[english]{babel}
\usepackage{graphicx}
\usepackage{amssymb}
\usepackage{amsmath}
\usepackage{overpic}

\newcommand{\be}{\begin{equation}}
\newcommand{\ee}{\end{equation}}
\newcommand{\bea}{\begin{eqnarray}}
\newcommand{\eea}{\end{eqnarray}}

\def\e{\varepsilon}

\def\l{\lambda}

\def\t{\tau}

\renewcommand{\o}{\omega}

\def\s{\sigma}

\def\D{\Delta}
\renewcommand{\O}{\Omega}

\def\ra{\rightarrow}
\def\up{\uparrow}

\def\down{\downarrow}

\def\pd{\partial}
\def\nb{\nabla}
\def\bk{{\bf k}}

\def\bq{{\bf q}}

\def\bA{{\bf A}}

\def\bE{{\bf E}}

\def\bJ{{\bf J}}

\def\nn{\nonumber}
\def\lb{\label}
\def\pref#1{(\ref{#1})}

\newcount\bozza \bozza=0
\ifnum\bozza=1
\newdimen\shift \shift=-2truecm
\def\lb#1{%
{\label{#1}\rlap{\kern\shift{$\scriptstyle#1$}}}}
\else\def\lb#1{\label{#1}} \fi

\begin{document}
\title{Non-linear optical effects and third-harmonic generation in superconductors: Cooper-pairs  vs Higgs mode contribution}
\author{T. Cea}
\affiliation{ISC-CNR and Dep. of Physics, ``Sapienza'' University of
  Rome, P.le A. Moro 5, 00185, Rome, Italy}
\email{lara.benfatto@roma1.infn.it}
\author{C. Castellani}
\affiliation{ISC-CNR and Dep. of Physics, ``Sapienza'' University of
  Rome, P.le A. Moro 5, 00185, Rome, Italy}
\author{L. Benfatto}
\affiliation{ISC-CNR and Dep. of Physics, ``Sapienza'' University of
  Rome, P.le A. Moro 5, 00185, Rome, Italy}
\date{\today}

\begin{abstract}
The recent observation of a transmitted Thz pulse oscillating at three times the frequency of the incident light paves the way to a new protocol to access resonant excitations in a superconductor. Here we show that this non-linear optical process is 
dominated by light-induced excitation of Cooper pairs, in analogy with a standard Raman experiment. The collective amplitude (Higgs) fluctuations of the superconducting order parameter give in general a smaller contribution, unless one designs the experiment by combining properly the light  polarization with the lattice symmetry.

\end{abstract}


\maketitle

The enormous technological advances made in the last two decades in the time-domain spectroscopy\cite{orenstein_review,giannetti_review} pose several challenges for our understanding of the interaction of the light with the matter. 
The use of low-energy THz waves\cite{review_thz} to first excite (pump) and then measure (probe) the system is particularly interesting for superconductors, since they can access the region $\omega<2\Delta_0$ of the optical spectrum where  linear-response absorption is suppressed by the opening 
of a superconducting (SC) gap $\Delta_0$ in the quasiparticle spectrum. For example, recent\cite{shimano_prl12,shimano_prl13,shimano14} THz pump-THz probe experiments have shown that the probe field displays a periodic oscillation, whose possible connection to amplitude (Higgs) fluctuations of the SC order parameter  has been investigated theoretically\cite{volkov73,levitov_prl04,levitov_prl06,axt_prb08,manske_prb14}.

\begin{figure}[t]
\includegraphics[width=8cm,clip=true]{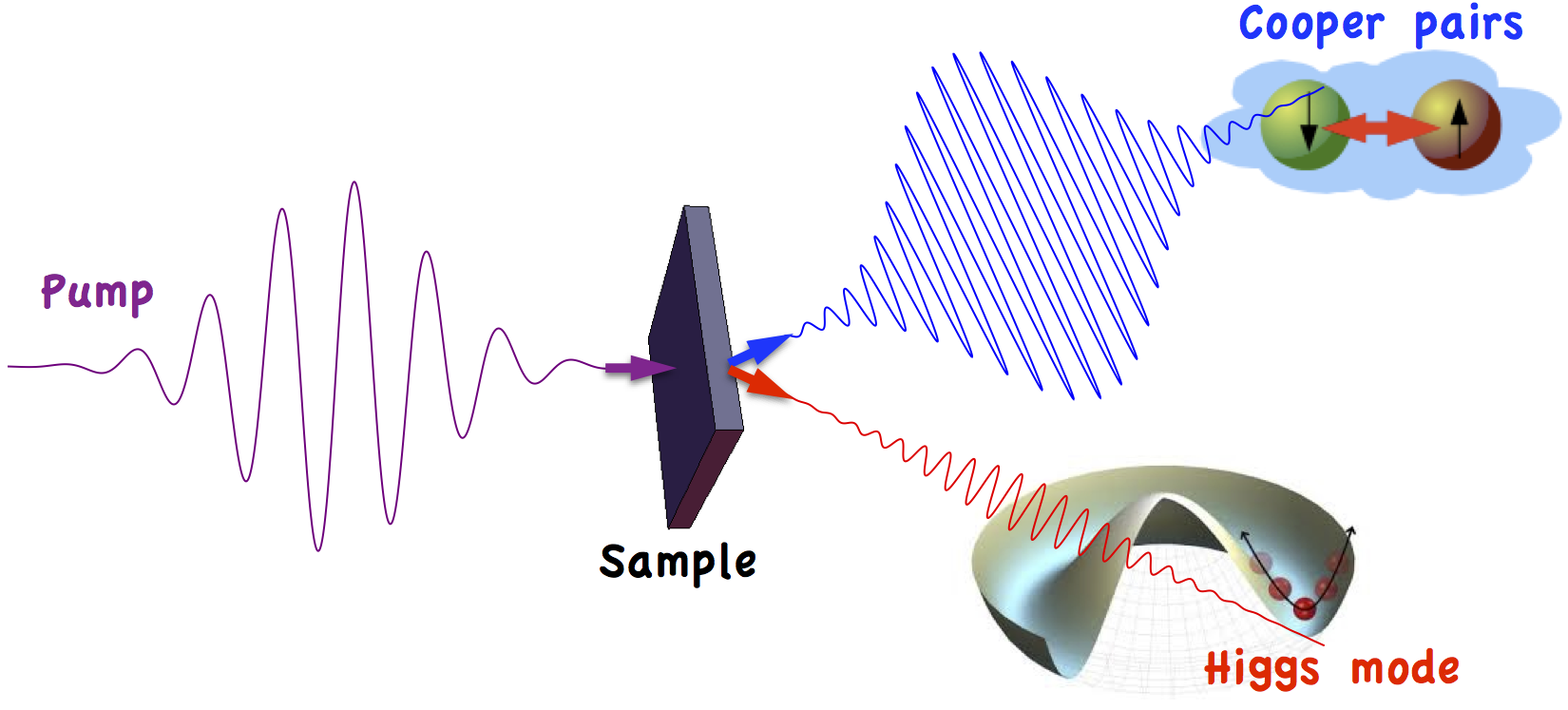}
\caption{Schematic of the THG. An intense  THz pulse shining on the SC sample generates a transmitted component  oscillating three times faster, due to the resonant excitations of Cooper pairs or Higgs fluctuations. The higher intensity of the former process can be modulated by changing the polarization of the incident light.}
\label{fig-schema}
\end{figure}

An interesting additional effect made possible by the use of intense electromagnetic (e.m.) THz field is the 
experimental observation\cite{shimano14} of the so-called third-harmonic generation (THG), i.e. the appearance below $T_c$ in the transmitted pulse of a component oscillating three times faster then the incident light. 
This effect appears only below $T_c$ with a maximum intensity at the temperature where the light frequency $\omega$ matches the SC gap value $\Delta_0(T)$, and has been attributed\cite{shimano14,TA} to a resonant excitation of the Higgs mode. However, we show here that THG is dominated by the resonant excitations of Cooper pairs (CP), (see Fig.\ \ref{fig-schema}), overlooked in previous theoretical work\cite{shimano14,TA}. 

In contrast to pump-probe experiments\cite{shimano_prl12,shimano_prl13}, 
where the description of the intermediate relaxation processes of the photoexcited states becomes relevant\cite{volkov73,levitov_prl04,levitov_prl06,axt_prb08,manske_prb14}, the THG effect can be understood as an equilibrium, non-linear  optical process. In this paper we compute microscopically  the non-linear optical response of a superconductor and we show that the THG essentially measures lattice-modulated {\em density} correlations, that in the SC state diverge at  the threshold $2\Delta_0$ above which Cooper pairs (CP) proliferate. This effect induces a resonant enhancement of the THG intensity when 
the frequency $2\omega$ of the incoming electric field coincides with $2\Delta_0$, as observed experimentally. Once identified the relevant non-linear optical response function, we also find that the Higgs-mode contribution is largely subleading, due to symmetry reasons. Indeed, even if the Higgs mode can be excited by the THz field, as discussed  previously\cite{shimano14,TA}, it essentially decouples from the optical probe. This is a consequence of the weak coupling between the SC amplitude and density fluctuations in BCS superconductors \cite{varma_prb82,cea_cdw_prb14,varma_review,cea_prl15}, as usually discussed in the context of Raman experiments\cite{hackl_review,varma_review,cea_cdw_prb14}. The potential analogy with Raman experiments emerges also on the non-trivial  dependence of the THG on the the relative orientation between the e.m. field and the main crystallographic axes, due to  the lattice symmetries of the band structure.    This effect can be tested e.g. in cuprate superconductors, where  large monocrystals have been already studied by non-linear spectroscopy\cite{carbone}.  Even though this polarization dependence can also be used to selectively excite the Higgs mode, its weak signal remains a major  obstacle to its detection.   Finally, by including the CP effects, missing in previous theoretical work\cite{shimano14,TA} due to an incorrect computation of the non-linear optical response, we reproduce very well the temperature dependence of the THG measured in Ref.\ \cite{shimano14}.

We start from a microscopic SC model that captures the main ingredients of the problem:
\be
\lb{hmodel}
H=\sum_{\mathbf{k},\sigma}\xi_\mathbf{k}c^\dagger_{\mathbf{k}\sigma}c_{\mathbf{k}\sigma}-\frac{U}{N_s}\sum_\bq \Phi^\dagger_\D(\bq)\Phi_\D(\bq),
\ee
where $\xi_\bk=\e_\bk-\mu$ is the electronic dispersion with respect to the chemical potential $\mu$, $U>0$ is the SC coupling and $\Phi_\Delta(\bq)=\sum_{\bk}c_{-\mathbf{k}+\mathbf{q}/2\downarrow}c_{\mathbf{k}+\mathbf{q}/2\uparrow}$.
 In mean-field approximation the Green's function in the usual basis of Nambu operators $\Psi^\dagger=(c^\dagger_{\bk\up}, c_{-\bk\down})$ reads $G_0^{-1}=i\o_n\hat\t_0-\xi_\bk\hat\t_3+\D_0\hat\t_1$, where $\hat\t_i$ are Pauli matrices,  $\D_0$ is the SC gap 
and $E_\bk=\sqrt{\xi_\bk^2+\D_0^2}$. The coupling to the 
gauge field $\bA$ can be introduced by means of the Peierls substitution, $c^\dagger_{i+\hat x}c_i\rightarrow c^\dagger_{i+\hat x}c_ie^{ie \bA\cdot \hat x}$. 
To derive the e.m. kernel we follow a standard procedure\cite{nagaosa,suppl} to derive  the action $S_A$ written in terms of the gauge field $\bA$ and SC collective modes. Since the coefficients of the effective action are given by fermionic susceptibilities this approach allows one to include both the quasiparticles and collective-mode contributions to the optical kernel, as it has been proven already for the linear response\cite{randeria_prb00,benfatto_prb04}. As we shall see below, it turns out that the most relevant contributions to the  non-linear current $\bJ^{NL}$ can be written in a compact notation as
\bea
\lb{jnlgen}
J^{NL}\sim  \bA (\chi^{CP}+\chi^H) \bA^2,\\
 \chi^{CP}\sim \langle \rho \rho \rangle, \quad
  \chi^{H}\sim  \langle \D \D\rangle.
\eea
where the CP contribution $\chi^{CP}$ probes lattice-modulated density fluctuations, while the Higgs contribution $\chi^H$ is proportional to the amplitude fluctuations. Even though both terms diverge at $\omega=\Delta_0$, the prefactor of $\chi^{H}$ turns out to be strongly suppressed by the particle-hole symmetry of the BCS solution, making it largely subdominant with respect to the CP one. 

To make this argument quantitative we compute the non-linear response by expanding the action $S_A$ up to the fourth order in $\bA$.
For an uniform field  the terms relevant for the THG are then:
\bea
\lb{SA}
S[A]&=&\frac{1}{2}\sum_{\O_n}  e^4 A_i^2(\O_n)\chi^{CP}_{ij}(\O_n)A_j^2(\O_n)+X_{\D\D}(\O_n)|\Delta(\O_n)|^2 \nn\\
&+&2e^2A_i^2(\O_n)\chi_{A^2_i\D}(\O_n)\Delta(-\O_n)
\eea
where $A_j^2(\O_n)$ is the Fourier transform of  $(A_j(t))^2$  in Matsubara frequency  $i\O_n=2\pi nT$.  The first term of Eq.\ \pref{SA} is the CP response, as given by
\bea
\lb{chi_0}
\chi^{CP}_{ij}(i\Omega_n)&=&\langle \rho_i \rho_j\rangle ={\Delta_0^2}\sum_{\mathbf{k}}\pd^2_i\e_\bk\pd^2_j\e_\bk F_\bk(i\O_n),\\
\lb{rk}
F_\bk(i\O_n)&=&\frac{1}{N_s}
\frac{\tanh(E_\mathbf{k}/2T)}{E_\mathbf{k}\left[(i\Omega_n)^2-4E_\mathbf{k}^2\right]},
\eea
where we introduced the short notation $\pd^2_i\e_\bk\equiv \pd^2 \e_\bk/\pd k^2_i$. Here $\langle \dots \rangle$ denotes the correlation function for the operator $\rho_i(\bq)=\sum_\bk \pd^2_i\e_\bk\ c^\dagger_{\bk+\bq}c_\bk$, showing that $\chi^{CP}_{ij}$ scales as the density-density correlation function (given in the BCS limit by   $\chi_{\rho\rho}\equiv\Delta_0^2\sum_\bk F_\bk(i\O_n)$), as anticipated in Eq.\ \pref{jnlgen} above.  Indeed, the band derivatives $\pd^2_i\e_\bk$ just represent in a lattice model the equivalent of the inverse mass $1/m$ for free electrons, and they always come along with a $A_i^2(\omega)$ term in the effective action\cite{benfatto_prb04,suppl}. The second term in Eq.\ \pref{SA} describes the collective fluctuations of the SC amplitude $\Delta$, $\langle |\Delta|^2\rangle_{\bA=0}=1/X_{\D\D}$, given as usual\cite{volkov73,kulik81,varma_prb82,cea_cdw_prb14,cea_prl15} by
\bea
\lb{xdd}
X_{\D\D}(\o)&=&(4\D_0^2-\o^2) F(\o),\quad F(\o)=\sum_\bk F_\bk(\o)
\eea
By analytical continuation $i\O_n\ra \o+i0^+$ one can easily see that both $\chi^{CP}(\o)$ and $F(\o)$ display a 
square-root divergence as $\o\ra 2\D_0$, that signals the proliferation of CP above the gap. As it is well known, this effects makes the Higgs a non-relativistic mode\cite{cea_prl15}, i.e. amplitude fluctuations display a overdamped  resonance at $\omega=2\Delta_0$. Finally, the third term in Eq.\ \pref{SA} describes the coupling between the e.m. field and the Higgs mode,  mediated  by the function
 \bea
 \lb{chi_A^2H}
\chi_{A_i^2\D}(i\Omega_n)
&=&	\langle \rho_i \Delta\rangle ={2\Delta_0}\sum_{\mathbf{k}}(\pd^2_i\e_\bk) \xi_\bk F_\bk(i\O_n).
\eea
Eq.s \pref{chi_0} and \pref{chi_A^2H} define the basic correlation functions needed to compute the THG. They also explain why for BCS superconductors, where they are given explicitly by the r.h.s. of Eq.\ \pref{chi_0}-\pref{chi_A^2H}, $\chi_{A_i^2\D}$ is very small. Indeed, in the continuum limit,  where the band dispersion can be approximated with a parabolic one $\e_\bk\simeq \bk^2/2m$,  so that $\pd_i^2 \e_\bk\simeq 1/m$, $\chi_{A^2_i\D}$ {\em vanishes}, since 
\bea
\chi_{A_i^2\D}(\o)\simeq  \frac{N_F}{m}\int_{-\infty}^\infty d\xi \frac{\xi}{\sqrt{\xi^2+\D_0^2}(\o^2-4\D_0^2-4\xi^2)}\simeq 0\nn,
\eea
where the integration range can be taken symmetric due to the approximate particle-hole symmetry of the BCS solution\cite{varma_prb82,cea_prl15}.  This result explains the suppression of the Higgs contribution to the THG. 

To derive the non-linear e.m. kernel we integrate out the amplitude fluctuations in Eq.\ \pref{SA}, that is equivalent to compute the RPA vertex correction of the bare bubble $\chi_{ij}^{CP}$\cite{TA,suppl}. One is then left with the action depending on the e.m. field only, 
\bea
\lb{S^4}
S^{(4)}[A]=\frac{e^4}{2}\int\,dt\,dt'\sum_{i,j}A_i^2(t) K_{ij}(t-t')A^2_j(t'),\\
\lb{kij}
K_{ij}(t-t')=\left[ \chi^{CP}_{ij}(t-t')+\chi^{H}_{ij}(t-t') \right]
\eea
where the Higgs contribution $\chi^H_{ij}$ reads
\begin{equation}
\lb{chi_nuc}
	\chi^{H}_{ij}(i\Omega_n)\equiv -\frac{\chi_{A_i^2\D}(i\Omega_n)\chi_{A_j^2\D}(i\Omega_n)}{X_{\D\D}(i\Omega_n)},
\end{equation}
and it is also diverging at $\o=2\Delta_0$ due to the vanishing of $X_{\D\D}$, see Eq.\ \pref{xdd}.
The non-linear current $J_i^{NL}$ follows by functional derivative  of Eq.\ \pref{S^4} with respect to $\bA$: 
\be
\lb{jnl}
J^{NL}_i(t)=-\frac{\delta S^{(4)}[A]}{\delta A_i(t)}=-2e^4A_i(t)\int\,dt'\sum_j K_{ij}(t-t')A^2_j(t'),
\ee
Eq.\ \pref{jnl}, with the definition \pref{kij} of the e.m. kernel, corresponds to Eq.\ \pref{jnlgen} above. For 
 a monocromatic incident field $\bA=\bar \bA\cos(\Omega t)$ it is given by
\bea
	J^{NL}_i(t)&=&\frac{e^4 \bar A_i}{4}\sum_j\left\{e^{-3i\Omega t}K_{ij}(2\Omega)
	+\right.\nn\\
\lb{jnlt}
&+&\left. e^{-i\Omega t}\left[ 2K_{ij}(0)+K_{ij}(2\Omega)\right]+c.c.\right\} \bar A_j^2
\eea
where one recovers the term oscillating at three times the incident frequency, with an amplitude controlled by the kernel $K_{ij}$ evaluated at $2\Omega$. 
In the experiments of Ref.\ [\onlinecite{shimano14}] the physical observable is the transmitted electric field $\bE^{tr}$, that one expects to be proportional to the current \pref{jnlt}. 
As a consequence the intensity of the THG can be evaluated from Eq.\ \pref{jnlt} as  $I^{THG}_{i}(\Omega)\propto \left|	\int\,dt J^{NL}_{i}(t)e^{3i\Omega t} \right |^2$, that for a monocromatic wave gives
\bea
\lb{ithg}
I^{THG}_{i}(\Omega)={I_0e^8 A_i^2}|\sum_j 	K_{ij}(2\Omega)\bar A_j^2	|^2,
\eea
where $I_0$ is an overall scale factor that depends on the geometry of the experiment. 
\begin{figure}[htb]
\includegraphics[width=7.5cm,clip=true]{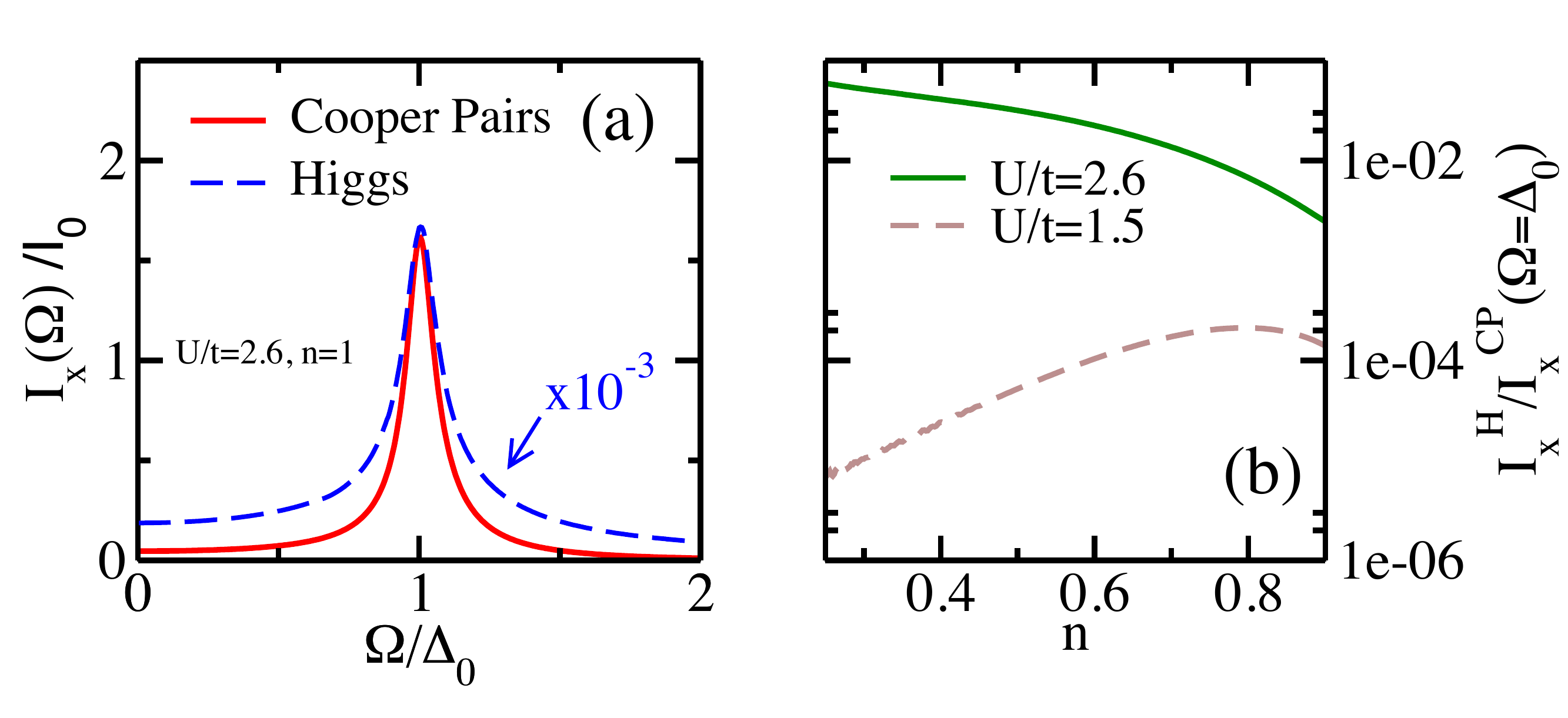}
\includegraphics[width=7.5cm,clip=true]{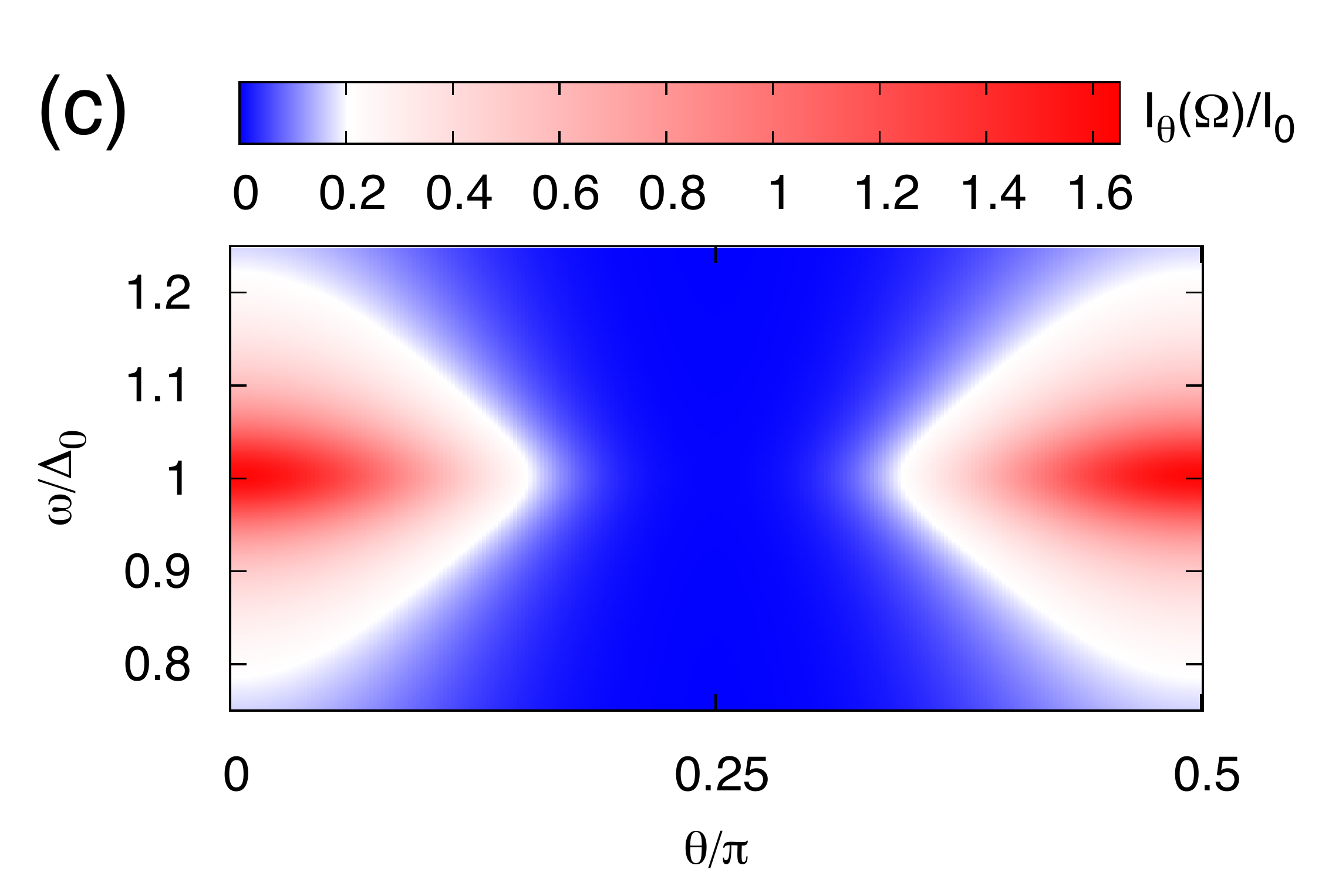}
\caption{(a) Comparison between the CP  and Higgs  contribution to the THG intensity at $T=0$ as a function of the frequency $\Omega$ for a field along $x$ and a residual broadening $\o+i\delta$, $\delta=0.1\Delta_0$. (b) Relative intensity of the Higgs and CP processes at $\omega=\D_0$  as a function of the density for two values of the SC coupling. Here the Coulomb screening from Eq.s \pref{chicpscr}-\pref{chiHscr} has been included. (c) Map of the frequency dependence of the THG intensity, Eq.\ \pref{Itheta}, for a field at arbitrary angle $\theta$ with respect to the $x$ direction. }
\label{fig-spectra}
\end{figure}

To quantify explicitly the lattice effects we compute the non-linear response for a nearest-neighbors tight-binding model on the square lattice $\e_\bk=-2t(\cos k_x+\cos k_y)$, and we will consider first the half-filled case $n=1$ ($\mu=0$), where only the SC amplitude mode contributes to the non-linear response.  By making the replacement $\pd^2_i\e_\bk=2t\cos k_i$ in Eqs. \pref{chi_0} and \pref{chi_A^2H} one sees that $\chi_{A^2\D}$ is independent on the direction while the 
CP part \pref{chi_0} is a tensor. Let us first consider the case of a field applied along the $x$ axis,  so that $I_x^{THG}$ is controlled by the longitudinal $K_{xx}$ kernel.  The two separate CP ($I^{CP}_x(\O)\propto|\chi^{CP}_{xx}(2\O)|^2$ or Higgs ($I^{H}_x(\O)\propto|\chi^{H}(2\O)|^2$ contributions to the THG intensity for a monocromatic field are shown in Fig.\ \ref{fig-spectra}a.   As one can see, even if the functional form is similar for the two terms, the CP contribution is much larger, and one can  roughly estimate $I^H_x\sim (\Delta/U)^4 I^{CP}_{xx}$. The predominance of the CP response implies also a non-trivial dependence of the THG intensity on the direction of the incoming applied field with respect to the crystallographic axes. In the general case of 
 $\bar \bA= A_0\cos(\O t)(\cos \theta,\sin \theta)$, $\theta$ being the angle with respect to the $x$ axis, the intensity of the transmitted pulse in the field direction is:
\bea
\lb{Itheta}
&&I^{THG}_\theta(\O)=I_0e^8A_0^6 |K_\theta(2\O)|^2\\
\lb{Ktheta}
&&K_\theta=\chi_{xx}^{CP}(\cos^4\theta+\sin^4\theta)+2\chi^{C}_{xy}\sin^2\theta\cos^2\theta+\chi^H
\eea
where we used the fact that $\chi_{yy/yx}^{CP}=\chi_{xx/xy}^{CP}$. When $\bar \bA$ is applied along the diagonal ($\theta=\pi/4$) the longitudinal $\chi_{xx}^{CP}$ and transverse $\chi_{xy}^{CP}$ parts of the CP response are equally weighted. In this peculiar configuration one sees from the definitions \pref{chi_0}-\pref{chi_A^2H} that  ${\chi^{CP}_{xx}+\chi^{CP}_{xy}}=-\Delta_0\chi_{A^2\Delta}/2$, i.e. the diverging CP contribution cancels out, and only the resonant Higgs response remains:
\begin{equation}
\lb{jpi4}
	J^{NL}_{\pi/4}(t)=\frac{e^2\Delta_0}{U}A(t) \langle\Delta(t)\rangle,
\end{equation}
where $\langle \D(t)\rangle$ is the average value of the amplitude fluctuations obtained from Eq.\ \pref{SA},  i.e. 
$\langle \Delta(\o)\rangle=e^2\chi_{A^2\Delta}(\o)\bA^2(\o)/X_{\D\D}(\o)$ in the frequency domain. The angular dependence of $I_\theta^{THG}(\O)$ for is shown in Fig.\ \ref{fig-spectra}c as a colour map: at $\theta=\pi/4$, where one probes only the Higgs mode, the intensity is strongly suppressed, in agreement with the result shown in Fig.\ \ref{fig-spectra}a. This prediction can be tested in systems like cuprate superconductors, where the band structure has in first approximation the symmetry discussed here and large monocristals have been already used to probe SC resonances by pumping the system with near-infrared light\cite{carbone}. It is worth noting that all these effects have been completely overlooked in the previous work\cite{shimano14,TA} due to the incorrect replacement in the CP term \pref{chi_0} of the quantity $\pd^2_i\e_\bk$, that is {\em finite} at the Fermi surface, with $\xi_\bk$, that is instead vanishing. This assumption  removes both the divergence of the CP term $\chi_{ij}^{CP}(\o)$ at $\o=2\Delta$ and its direction dependence, and leads always to the result \pref{jpi4}, that is instead far from being generic. In addition, we also checked\cite{suppl} that the expression \pref{kij} can be obtained as well by means of the  pseudospin formalism used in Ref.\ \cite{shimano14,TA}.

Since the response function \pref{kij} is dominated by the electronic states at the Fermi surface, the quantitative difference between the CP and Higgs response depends in general on the electron density $n$.  To quantify this effect away from half-filling one should retain in the derivation \pref{SA} of the effective action also the terms\cite{suppl} coupling the gauge field and the Higgs mode to the phase and density fluctuations, mediated by the response functions
\bea
\lb{chiA^2r}
\chi_{A_i^2\rho}(i\Omega_n)&=&\langle \rho_i \rho \rangle ={\Delta_0^2}\sum_\bk \pd^2_i\e_\bk F_\bk(i\O_n),\\
\lb{chiDr}
\chi_{\rho\Delta}(i\Omega_n)&=&\langle \rho\D\rangle ={\Delta_0}\sum_\bk \xi_\bk F_\bk(i\O_n).
\eea
These terms, that vanish by particle-hole symmetry in the half-filled case, are crucial to account for the screening effects of the long-range Coulomb potential, in analogy again with the known result for the Raman response function\cite{hackl_review}. 
By means of straightforward but lengthly calculations one can then show\cite{suppl} that the non-linear response function retains the structure \pref{kij} with the replacements
\bea
\lb{chicpscr}
	\chi^{CP}_{ij}&\rightarrow&  \chi^{CP,sc}_{ij}=	\chi^{CP}_{ij}-\frac{\chi_{A^2\rho}^2}{\chi_{\rho\rho}},\\
\lb{chiHscr}
	\chi^{H}&\rightarrow&\chi^{H,sc}=-\frac{\left(	\chi_{A^2\D}-\chi_{A^2\rho}\chi_{\rho \D}/\chi_{\rho\rho}\right)^2}{X_{\D\D}-\chi_{\rho \D}^2/\chi_{\rho\rho}}.
\eea
where we used the fact that for the lattice model under consideration the function \pref{chiA^2r} is isotropic in the spatial indexes. While the mixing to the density and phase modes does not affect\cite{cea_prl15} the pole of the Higgs,  identified now by the vanishing of the denominator of Eq.\ \pref{chiHscr}, it is crucial to screen both the CP and Higgs response as one moves away from half-filling. In Fig.\ \ref{fig-spectra}b  we show the ratio $I_{x}^{CP}(\D)/I_x^{H}(\D)$ as a function of the electron density $n$ for two values of the SC coupling $U$. As we can see, even for the large value $U/t=2.6$ of the SC coupling, where the deviations from the BCS (approximate) particle-hole symmetric case become more prominent, inducing a larger coupling of the Higgs to the light, the CP part remains the predominant one for the longitudinal response even in the low-density regime. 

\begin{figure}[t]
\includegraphics[width=9.5cm,clip=true]{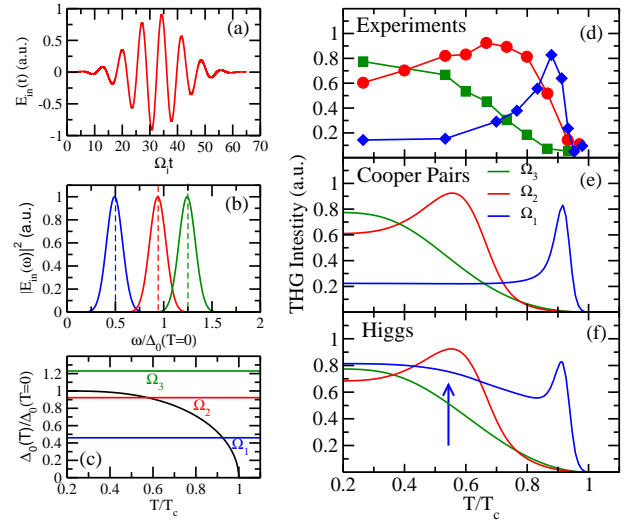}
\caption{(a) Profile of the incoming field, and (b) corresponding power spectra, used to simulate the experiments of Ref.\ \cite{shimano14}. The three central frequencies $\O_i$ of the power spectra are compared in panel (c) to the temperature dependence of the SC gap. For each $\O_i$ the resonant condition $\Omega_i=\Delta_0(T)$ occurs at a different temperature, in analogy with Ref.\ \cite{shimano14}. (d-e) 
Temperature evolution of the THG intensity in the experiments (d), in the case of CP processes (e) and in the case of the Higgs processes alone (f), computed for $n=1, U/t=2.6$. Data computed at different $\O_i$ are normalized to have a similar overall scale, as done in panel (d). The excitation of the Higgs mode alone gives the wrong $T$ dependence of the THG signal for the lowest frequency $\O_1$, as marked by the arrow.}
\label{fig-intensity}
\end{figure}
In addition to the strong direction dependence of the THG intensity, a second check of the origin of the THG effect is its temperature evolution, measured in Ref.\ [\onlinecite{shimano14}]. For a policristalline sample one should 
average the kernel Eq.\ \pref{Ktheta} over $\theta$, to account for the random direction of the e.m. field with respect to the crystallographic axes. One then finds that ${\bar J^{NL}_\theta}=\left[ J_{x}^{NL}+J_{\pi/4}^{NL}\right]/2\simeq J_{x}^{NL}/2$, so that one expects that the CP processes dominate. To check this we compute\cite{suppl} the non-linear current induced by an incoming electric field $\bA(t)$ having a wave-packet profile similar to the one used in the experiments of Ref.\ \cite{shimano14}, see Fig.\ \ref{fig-intensity}a. In the frequency domain this wave packet corresponds to the power spectra shown in Fig.\ \ref{fig-intensity}b, centred at three possible values $\O_i$ of the incoming frequency. The temperature evolution of the THG intensity, i.e. $I_x(3\Omega_i)$,  is then shown in Fig.\ \ref{fig-intensity}d-f, where we compare the experimental data from Ref.\ \cite{shimano14} (panel d) with the theoretical calculations done including only the CP processes (panel e) or the  Higgs contribution (panel f). Apart from the small overall intensity of the THG Higgs signal, that cannot be seen in the normalized data of  Fig.\ \ref{fig-intensity}, the excitation of the Higgs mode alone fails to reproduce the temperature dependence of the signal at the lowest frequency $\Omega_1$. 

In conclusion, we studied the non-linear optical effects responsible for the THG  in a superconductor.  Since the relevant response function \pref{kij} measures lattice-modulated density fluctuations, see Eq.\ \pref{chi_0},  the optical process responsible for the THG is equivalent to a resonant excitation of CP. The Higgs-mode contribution is instead much smaller, since its coupling \pref{chi_A^2H} to the optical probe is suppressed by symmetry, in analogy with 
standard Raman experiment\cite{hackl_review,varma_prb82, cea_cdw_prb14,varma_review}, unless  some additional channel makes the Higgs Raman visible\cite{varma_prb82,cea_cdw_prb14,measson_prb14}. In addition, also for THG experiments one can orient the light polarization with respect to the main crystallographic axes in order to modulate the THG intensity due to Cooper pairs. Even though this effect can be used in principle to selectively excite the Higgs signal, the weakness of its coupling to the light represents a major obstacle to its detection in optical experiments, in analogy with the results recently discussed in the context of linear optical spectroscopy\cite{cea_prl15}. The interplay between the non-linear optical effects discussed here and the non-equilibrium processes\cite{volkov73,levitov_prl04,levitov_prl06,axt_prb08,manske_prb14} addressed in the context of pump-probe experimental protocols\cite{orenstein_review,giannetti_review,shimano_prl12,shimano_prl13} remains an open question, that certainly deserves future experimental and theoretical work.


\vspace{0.5cm}  
  
This work has been supported  by Italian MIUR under projects FIRB-HybridNanoDev-RBFR1236VV, PRINRIDEIRON-2012X3YFZ2 and Premiali-2012 ABNANOTECH.

\vspace{1cm}

\pagebreak
\clearpage

\onecolumngrid
\begin{center}
\textbf{\Large Supplemental Material}
\end{center}
\vspace{1cm}
\twocolumngrid

\setcounter{equation}{0}
\setcounter{figure}{0}
\setcounter{table}{0}
\setcounter{page}{1}
\makeatletter
\renewcommand{\theequation}{S\arabic{equation}}
\renewcommand{\thefigure}{S\arabic{figure}}
\renewcommand{\bibnumfmt}[1]{[S#1]}
\renewcommand{\citenumfont}[1]{S#1}

\section{Derivation of the non-linear optical kernel}
Let us start from Eq. (3) of the manuscript, that can be rewritten in real space as:
\begin{equation}\lb{H_MODEL}
H=-t\sum_{\langle i,j\rangle\sigma}\left(c^\dagger_{i\sigma}c_{j\sigma}+h.c.\right)-U\sum_ic^\dagger_{i\uparrow}c^\dagger_{i\downarrow}c_{i\downarrow}c_{i\uparrow}.
\end{equation}
To investigate the physics of the collective fluctuations around the mean field solution we follow the usual Hubbard-Stratonovich (HS) procedure\cite{nagaosa}, as implemented e.g. in Refs. \onlinecite{depalo_prb99,benfatto_prb04}. We then introduce in the action for the fermions a bosonic complex field $\psi_\D(\tau)$ which decouples the onsite interaction term of \eqref{H_MODEL} in the pairing channel. At $T<T_c$ one can choose to represent the superconducting (SC) fluctuations  in polar (amplitude and phase)  coordinates, by decomposing $\psi_\D(\tau)=[\Delta_0 +\Delta_i(\tau)]e^{i\theta_i(\tau)}$, where $\Delta_i(\tau)$ represent the amplitude fluctuations of $\psi_\D$ around the mean-field value $\Delta_0$ of the SC order parameter and $\theta$ its phase fluctuations. By making a Gauge transformation $c_i\ra c_ie^{i\theta_i/2}$ the dependence on the phase degrees of freedom is made explicit in the action. We also add to the model \pref{H_MODEL} an additional interaction term representing the effect of Coulomb interactions:
\be
\lb{hc}
H_c=\frac{1}{2}\sum_{\bk,\bk',\bq\atop{\sigma\sigma'}} V(\bq)c^\dagger_{\bk+\bq,\s}c^\dagger_{\bk'-\bq,\s'}c_{\bk',\s'}c_{\bk,\s}
\ee
where $V(\bq)$ is the Fourier transform of the Coulomb potential in the $D$ dimensional lattice. At small $\bq$ it reduces to the expression in the continuum limit, so that $V(\bq)\ra\lambda e^2/|\bq|^{D-1}$ where $\lambda=4\pi/\epsilon_B$ for $D=3$ while $\l=2\pi/\epsilon_B$ for $D=2$, $\e_B$ being the background dielectric constant. This term can be decoupled in the particle-hole channel by means of an additional  HS field $\psi_\rho=\rho_o+\rho$, which couples to the electronic density $\Phi_{\rho,i}=\sum_\sigma c^\dagger_{i\sigma}c_{i\sigma}$ and represents the density fluctuations $\rho$ of the system around the mean-field value $\rho_0$. Finally, the Gauge field $\bA$ can be introduced by means of the Peiersl substitution  $c^\dagger_{i+\hat x}c_i\rightarrow c^\dagger_{i+\hat x}c_ie^{ie \bA\cdot \hat x}$, that modifies only the kinetic part of the Hamiltonian. 

After the Hubbard-Stratonovich decoupling the action is quadratic in the fermionic fields that can then be integrated out leading to the effective action for the fields $\Delta$, $\theta$ , $\rho$ and $\bA$:
\begin{equation}
\lb{seff}
	S_{eff}[\Delta,\theta,\rho]=S_{MF}+S_{FL}[\Delta,\theta,\rho,\bA]\quad,
\end{equation}
where $S_{MF}=\frac{N\Delta_0^2}{TU}+\frac{N\rho_0^2}{TU}-\text{Tr}\ln(-G_0^{-1})$ is the mean-field action, 
$G_0^{-1}=i\o_n\hat\s_0-\xi_\bk\hat\s_3+\D_0\hat\s_1$ is the BCS Green's function and 
\be 
\lb{sfl}
S_{FL}=\sum_{n\ge1}\frac{\text{Tr}(G_0\Sigma)^n}{n}
\ee
is the fluctuating action, with the trace acting both in spin and momentum space. Here $\Sigma_{kk'}$ denotes the self-energy for the fluctuating fields, which reads explicitly:
\begin{widetext}
\bea
\Sigma_{kk'}&=&
-\sqrt{\frac{T}{N}}\Delta(k-k')\sigma_1-\sqrt{\frac{T}{N}}\rho(k-k')\sigma_3-
\sqrt{\frac{T}{N}}\frac{i}{2}\theta(k-k')\left[
(k-k')_0\sigma_3-(\xi_\bk-\xi_{\bk'})\sigma_0
\right]-\nn\\
&-&\frac{T}{2N}\sum_{q_i,i}\theta(q_1)\theta(q_2) \frac{\pd^2 \xi_\bk}{\pd k^2_i} \sin\frac{\bq_{1,i}}{2}\sin\frac{\bq_{2,i}}{2}
\sigma_3 \delta(q_1+q_2-k+k')+\nn\\
\lb{SELF_ENERGY}
&+&\sum_i \left[A_i(\o-\o')\frac{\pd \xi_\bk}{\pd k_i}\sigma_0+\frac{1}{2}A_i^2(\o)\frac{\pd^2 \xi_\bk}{\pd k^2_i}\sigma_3+
\frac{1}{3!}A_i^3(\o)\frac{\pd^3 \xi_\bk}{\pd k^3_i}\sigma_0+\frac{1}{4!}A_i^4(\o)\frac{\pd^4 \xi_\bk}{\pd k^4_i}\sigma_3\right]
\eea
\end{widetext}
with $k=(i\Omega_n,\mathbf{k})$ and $\Omega_n=2\pi Tn$ bosonic Matsubara frequencies. Here 
$A_j^2(\omega)\equiv \int\,d\omega' A_j(\omega-\omega')A_j(\omega')$ is the Fourier transform of $\left[A(t)\right]^2$, and analogous convolution formulae hold for higher powers of the gauge field. The second line of Eq.\ \pref{SELF_ENERGY} represents the transcription on the lattice of the usual $(\nb \theta)^2$ term for a continuum model, analogously to the $A_i^2(\o)$ term that represents the transcription of the usual diamagnetic term $\bA^2 n/m$ in the continuum. In addition, in contrast to the continuum model, the lattice self-energy \pref{SELF_ENERGY} depends in principle\cite{depalo_prb99,benfatto_prb04} on all higher-order powers of the $\theta$ and $\bA$ fields. Here however we only retained the terms relevant for the derivation of the action up to terms of order $A_i^4$, and we considered directly the case of an uniform e.m. field.

\begin{figure}
\centering
\includegraphics[scale=0.45]{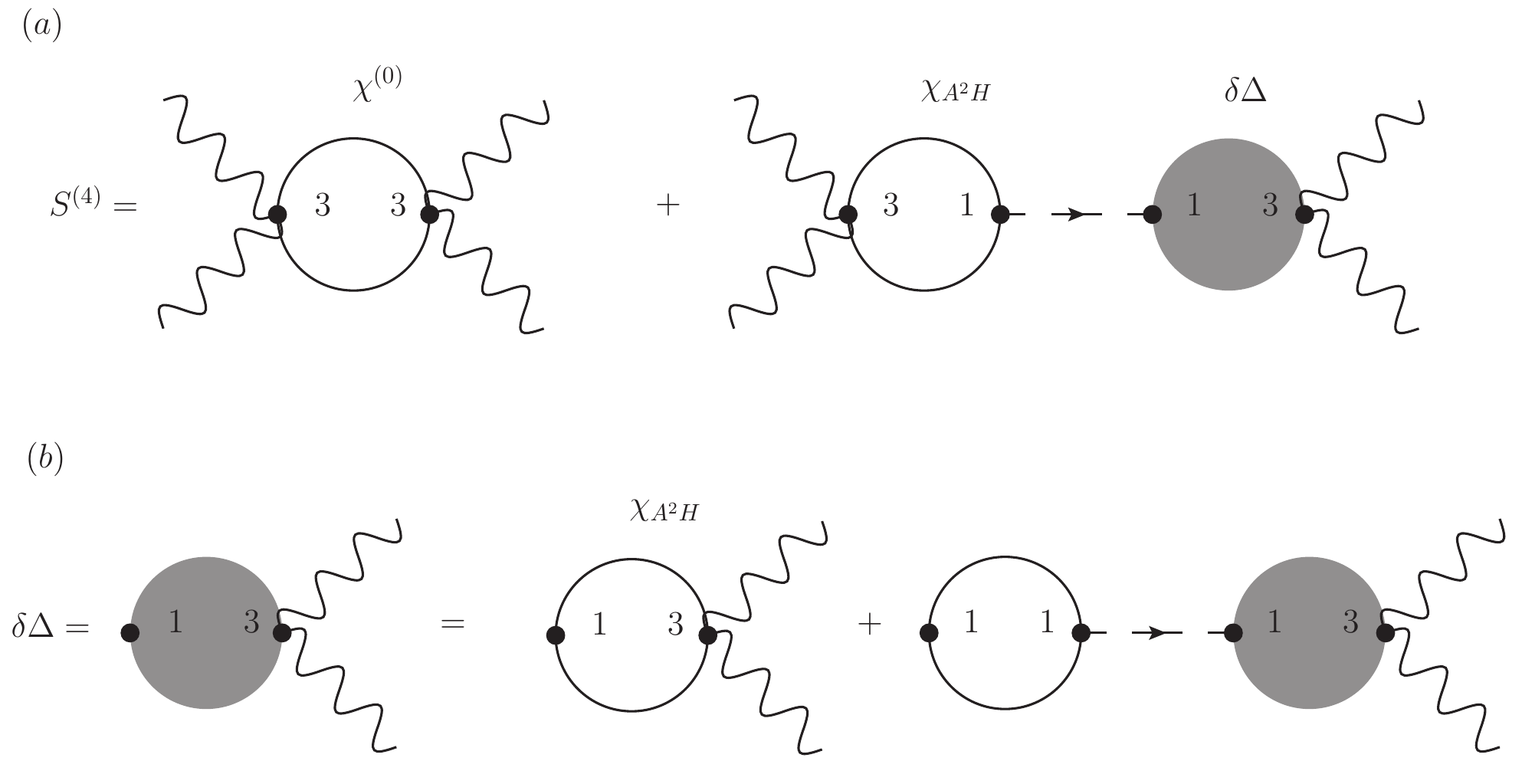}
\caption{
Figure (a): diagrammatic representation of $S^{(4)}$. The wavy and dashed lines denote the gauge and Higgs fields, respectively, while the labels $1$, $3$ refer to the vertex insertions of the Pauli matrices $\sigma_1$ and $\sigma_3$, respectively. Figure (b) represents the vertex correction of $\chi_{A^2\D}$ in the amplitude channel, which defines the variation of the order parameter, $\delta\Delta$, due to the applied external field $\mathbf{A}$.
}\label{Diagrams}
\end{figure}

As one can see from Eq.\ \pref{sfl} the effective action gives an expansion on powers of the bosonic fields, whose coefficients are fermionic susceptibilities that contain all the relevant information on the quasiparticle degrees of freedom. Away from half-filling the phase/density fluctuations are in general coupled both to the gauge field and to the amplitude fluctuations $\D$. The expansion of $S$ up to quartic order in $\bA$ is then given by the generalization of Eq.\ (5) of the manuscript, i.e.
\bea
S[A]&=&\frac{1}{2}\sum_{\O_n}  e^4 A_i^2(\O_n)\chi^{CP}_{ij}(\O_n)A_j^2(\O_n)+\nn\\
&+&2e^2A_i^2(\O_n)\chi_{A^2_i\D}(\O_n)\Delta(-\O_n)+\nn\\
&+&2e^2A_i^2(\O_n)\chi_{A^2_i\rho}(\O_n)\left[\rho(-\O_n)-i\O_n\theta(-\O_n)\right]+\nn\\
\lb{SA}
&+&\frac{1}{2}\sum_q
\Psi^\dagger(q)
	\hat{M}_{FL}(q)
	\Psi(q)\quad,
\eea
where  $\Psi^T(q)=\begin{pmatrix}\Delta(q)& \theta(q) & \rho(q)\end{pmatrix}$ is a vector containing the fluctuating fields, whose Gaussian fluctuations are described by the matrix\cite{cea_prl15}:
\begin{equation}
\lb{matrix}
\hat{M}_{FL}=\begin{pmatrix}
	2/U+\chi_{\D\D}(\O_n)
	&
	\frac{i \O_n}{2}\chi_{\rho\D}(\O_n)&
	\chi_{\rho\D}(\O_n)\\
	-\frac{i\O_n}{2}\chi_{\rho\D}(-\O_n)&
	\frac{\O_n^2}{4}\chi_{\rho\rho}(\O_n)&
	-	\frac{i\O_n}{2}\chi_{\rho\rho}(\O_n)\\
		\chi_{\rho\D}(-\O_n)&
			\frac{i\O_n}{2}\chi_{\rho\rho}(\O_n)&
				-1/V_\bq+\chi_{\rho\rho}(\O_n)
	\end{pmatrix}.
\end{equation}
The various bubbles, given explicitly in the main manuscript,  are defined in terms of the Green's functions, as
\begin{widetext}
\begin{equation}\label{chi_0}
	\chi^{CP}_{ij}(i\Omega_n)\equiv\frac{T}{N_s}\sum_{\mathbf{k},m}\pd^2_i\e_\bk\pd^2_j\e_\bk \text{Tr}\left[G_0(\mathbf{k},i\omega_m+i\Omega_n)\sigma_3G_0(\mathbf{k},i\omega_m)\sigma_3\right]
\end{equation}
\begin{equation}\label{chi_A^2H}
	\chi_{A_i^2\D}(i\Omega_n)\equiv \frac{T}{N_s}\sum_{\mathbf{k},m}\pd^2_i\e_\bk\text{Tr}\left[G_0(\mathbf{k},i\omega_m+i\Omega_n)\sigma_3G_0(\mathbf{k},i\omega_m)\sigma_1\right]
\end{equation}
\begin{equation}\label{chi_A^2r}
	\chi_{A_i^2\rho}(i\Omega_n)\equiv \frac{T}{N_s}\sum_{\mathbf{k},m}\pd^2_i\e_\bk\text{Tr}\left[G_0(\mathbf{k},i\omega_m+i\Omega_n)\sigma_3G_0(\mathbf{k},i\omega_m)\sigma_3\right]
\end{equation}
\begin{equation}
	\chi_{\rho\D}(i\Omega_n)\equiv \frac{T}{N_s}\sum_{\mathbf{k},m}\text{Tr}\left[G_0(\mathbf{k},i\omega_m+i\Omega_n)\sigma_3G_0(\mathbf{k},i\omega_m)\sigma_1\right]
\end{equation}
\begin{equation}
	\chi_{\rho\rho}(i\Omega_n)\equiv \frac{T}{N_s}\sum_{\mathbf{k},m}\text{Tr}\left[G_0(\mathbf{k},i\omega_m+i\Omega_n)\sigma_3G_0(\mathbf{k},i\omega_m)\sigma_3\right]
\end{equation}

\begin{equation}\label{X_HH}
	X_{\D\D}(i\Omega_n)\equiv\frac{2}{U}+\chi_{\D\D}=\frac{2}{U}+ \frac{T}{N_s}\sum_{\mathbf{k},m}\text{Tr}\left[G_0(\mathbf{k},i\omega_m+i\Omega_n)\sigma_1G_0(\mathbf{k},i\omega_m)\sigma_1\right]\end{equation}

\end{widetext}
As one can see, apart from the modulation factors $\pd^2_i\e_\bk$, one recovers that $\chi_{A^2_i\D}\sim \chi_{\rho\D}$ and $\chi_{A^2_i\rho}\sim \chi_{\rho\rho}$. While in the low-density limits the derivatives of the band dispersions just reduces to a constant, at half-filling ($\mu=0$), where particle-hole symmetry is exact, they are crucial to decouple the amplitude mode and the e.m. field from the density/phase modes. In this case one only retains the second line of Eq.\ \pref{SA}, and after integration of the Higgs mode one is left with Eq.\ (10) of the manuscript,
\bea
\lb{S^4}
S^{(4)}[A]=\frac{e^4}{2}\int\,dt\,dt'\sum_{i,j}A_i^2(t) K_{ij}(t-t')A^2_j(t'),\\
\lb{kij}
K_{ij}(t-t')=\left[ \chi^{CP}_{ij}(t-t')+\chi^{H}_{ij}(t-t') \right],
\eea
where 
\begin{equation}\label{chi_nuc}
	\chi^{H}_{ij}(i\Omega_n)\equiv -\frac{\chi_{A_i^2\D}(i\Omega_n)\chi_{A_j^2\D}(i\Omega_n)}{X_{\D\D}(i\Omega_n)}\quad,
\end{equation}
The structure of the quartic action \eqref{S^4} is diagrammatically represented in fig. \ref{Diagrams}. Here $\delta\Delta$ denotes the variation of the order parameter from its equilibrium value due to the external perturbation, which is obtained by dressing $\chi_{A^2H}$ with the vertex correction in the amplitude channel (see fig. \ref{Diagrams}-(b)):
\begin{equation}\lb{deltaDELTA}
\delta\Delta(\omega)=e^2\frac{\chi_{A^2\D}(\omega)}{X_{\D\D}(\omega)} \bA^2(\o).
\end{equation}
\begin{figure}
\centering
\includegraphics[scale=0.3]{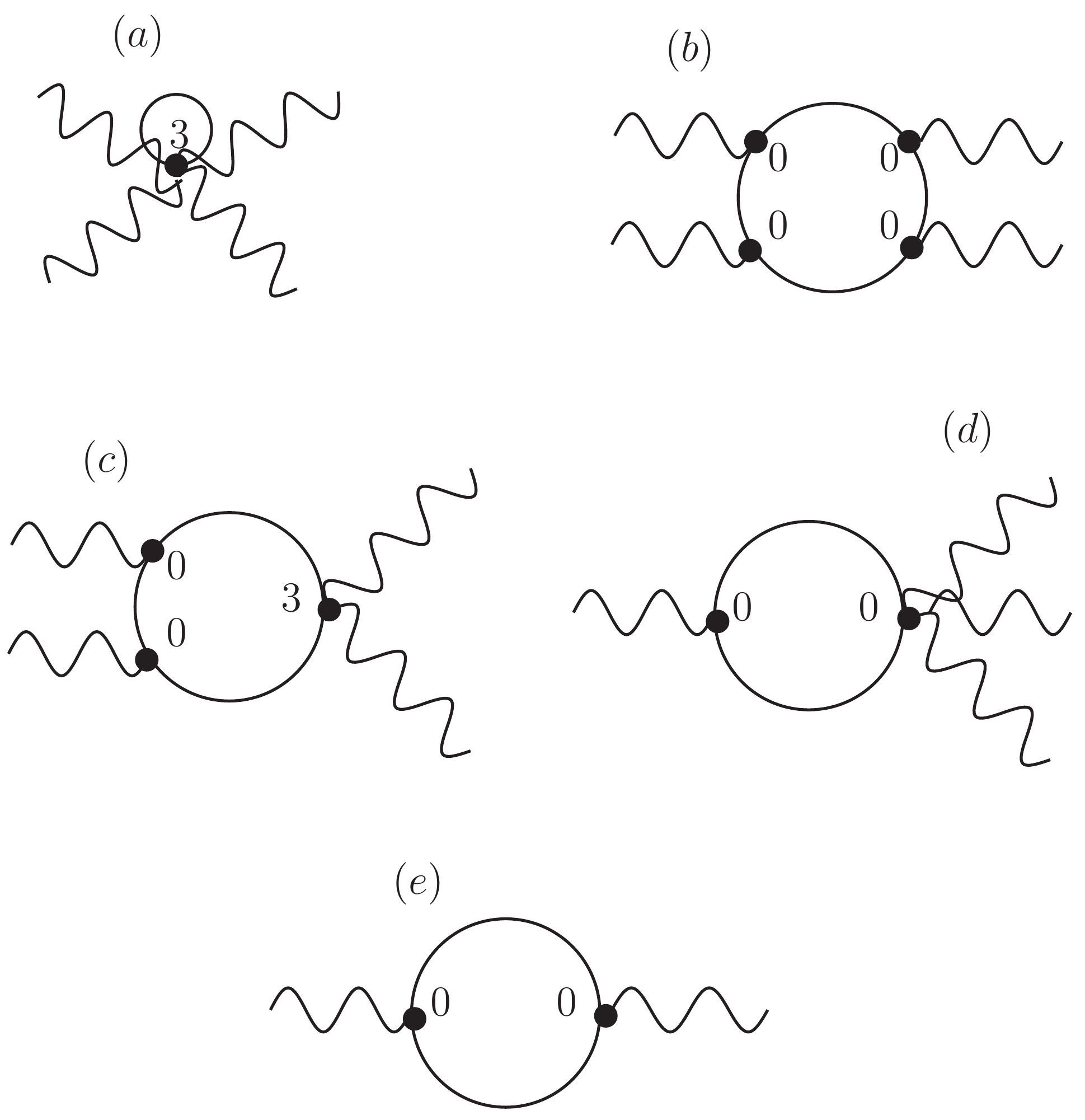}
\caption{
(a)-(d) Additional diagrams contributing to $S^{(4)}$. The (a)  is independent on the transferred frequencies, while the remaining ones (b)-(d) are trivially zero when computed at zero external momenta. (e) Diagram defining the paramagnetic response in the linear response regime. It leads to the emptying of the condensate in the static limit ($\omega=0$, $\mathbf{q}\to0$), whereas it vanishes at $\mathbf{q}=0$ and finite $\omega$.
}\label{Diagrams2}
\end{figure}

Notice that in addition to the diagrams shown in Fig.\ \ref{Diagrams}, and included in the Eq.\ \pref{S^4}, one can have in principle several other terms of order $A_i^4$, coming from the insertion of various $A_i^n$ term of the self-energy \pref{SELF_ENERGY},  as shown in Fig.\  \ref{Diagrams2} (a)-(d). They have been omitted in $S^{(4)}$ since the first one is independent on the transmitted frequencies, while those having the $\sigma_0$ insertions trivially vanish when computed at zero external momenta. This is indeed a general result which follows from elementary algebra principles and holds for the whole class of diagrams having an arbitrary number of insertions of $\sigma_0$ and only one insertion of $\sigma_i$ (with $i=0,\dots3$). Despite such a rule does not hold any more in the presence of impurities, we expect that all this kind of diagrams will still be regular functions of the external frequencies. An illustrative example is represented by the diagram \ref{Diagrams2}-(e), which defines the paramagnetic response in the linear response regime. As it is well known\cite{benfatto_prb04}, in the static limit ($\omega=0$, $\mathbf{q}\to0$) it accounts for the reduction of the superfluid stiffness at finite temperature due to thermal excitations of quasiparticles, while it vanishes in the opposite dynamic limit ($\mathbf{q}=0, \omega\neq 0$). In the presence of disorder it also contributes to the dynamic limit, leading to the optical absorption at $\omega>2\Delta_0$, without however any divergence at $\omega=2\Delta_0$, in contrast to the diagrams constructed with two $\sigma_3$ insertions.

Away from half-filling the  phase/density fluctuations couple to the fields $\D$ and $A$. In particular, the density fluctuations $\rho$, mediating the Coulomb repulsion, screen the other fields. By expliciting integrating the density and phase fields from Eq.\ \pref{SA}, and taking the lmit $\bq\ra 0$ where $1/V_\bq\ra 0$, one is left with the same structure \pref{kij} of the non-linear kernel, provided that the CP and Higgs part as screened as given by Eqs. (19)-(20) of the main text, i.e.
\bea
\lb{chicpscr}
	\chi^{CP}_{ij}&\rightarrow&  \chi^{CP,sc}_{ij}=	\chi^{CP}_{ij}-\frac{\chi_{A^2\rho}^2}{\chi_{\rho\rho}},\\
\lb{chiHscr}
	\chi^{H}&\rightarrow&\chi^{H,sc}=-\frac{\left(	\chi_{A^2\D}-\chi_{A^2\rho}\chi_{\rho \D}/\chi_{\rho\rho}\right)^2}{X_{\D\D}-\chi_{\rho \D}^2/\chi_{\rho\rho}}.
\eea
%

\section{THG by an impulsive electric field}

In the previous section we have studied the THG when the applied electric field is a monochromatic wave. To better simulate the experiment described in \cite{shimano14} we now consider an impulsive multi-cyclic  field, having a relatively narrow spectrum of frequencies. To this purpose we choose:
\begin{equation}
\lb{a0}
	A(t)=A_0e^{-(t/\tau)^2}\cos\left(\Omega t\right)\quad,
\end{equation}
where $\tau$ is a time constant which determines the duration of the impulse. The incoming electric field, given by the time derivative of Eq.\ \pref{a0}, is shown in Fig. 3a of the manuscript, and its corresponding power spectrum is shown in Fig.\ 3b for three different values of the central frequency $\O=\O_i$. By using the definition of the non-linear current given in Eq. (14) of the main text one easily finds that:
\begin{widetext}
\begin{equation}
	J^{NL}_{\theta}(\omega)=
	-\frac{e^4A_0^3\tau^2}{\sqrt{8}}\int\,d\omega' K_{\theta}(\omega')\exp\left[	-\frac{\tau^2}{8}\left(3\omega'^2+2\omega^2-4\omega\omega'+2\Omega^2	\right)\right]
	\cosh\left[\frac{\tau^2}{2}\Omega\left( \omega-\omega' \right)\right]
	\left[	1+e^{-\frac{\tau^2}{2}\Omega^2}\cosh\left(\frac{\tau^2}{2}\Omega\omega'\right)	\right]
	\text{ }.
\end{equation}
\end{widetext}

\begin{figure}[h]
\centering
\includegraphics[scale=0.33]{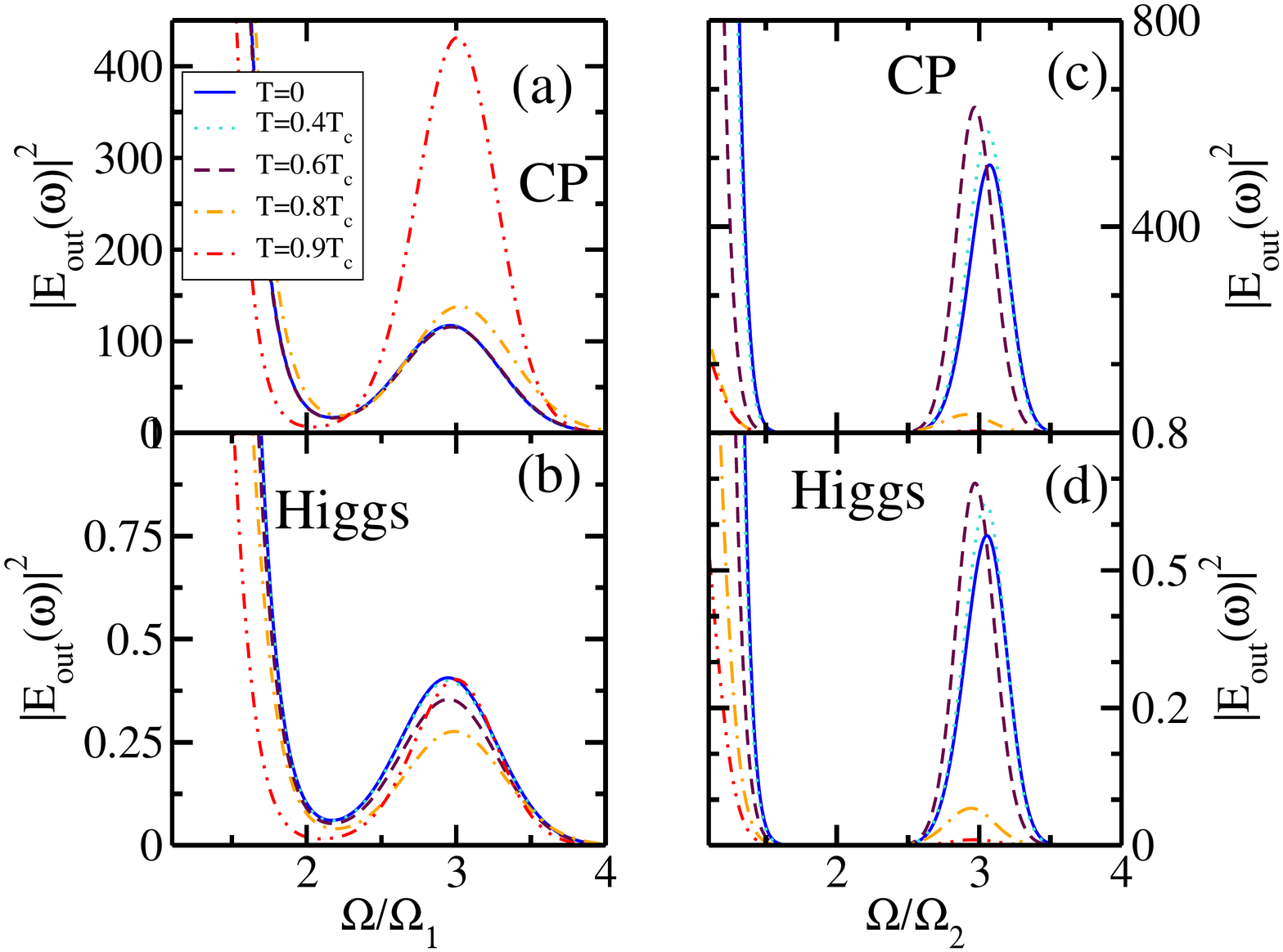}
\caption{Temperature evolution of the power spectra of the outgoing electric field generated by CP (a,c) or Higgs (b,d) processes corresponding to the incoming THz pulse \pref{a0} for two different values of the central frequency $\O$. The peak around $3\O_i$ defines the THG intensity. The calculations have been done in the case $n=1, U/t=2.6$. Simulations at different electron densities give similar results for the relative $T$ variations.}
\label{fig-power}
\end{figure}

As we discuss in the main text, the outgoing electric field is expected to be proportional to the non-linear current, $|E_{out}(\o)|^2\propto |J^{NL}_x(\o)|^2$. The power spectra of the outgoing fields generated by the CP or Higgs processes are shown in Fig.\ \ref{fig-power} for the lowest ($\O_1$) and intermediate ($\O_2$) value of the central frequency. As one can see, in addition to the peak at $\o=\O_i$ an additional peak appears around $\o=3\O_i$, whose absolute intensity and temperature variation depend on the nature of the resonant process. Notice the close resemblance between the experimental data shown in Fig. 3D of Ref.\ \cite{shimano14} and  the temperature evolution of the power spectra for $\O=\O_2$, where the variation of THG intensity is accompanied by a small shift of the peak maximum. To make a direct comparison with the data shown in Ref.\ \cite{shimano14}, we adopt the same definition of THG intensity at each temperature as the one given there, i.e. we take the value at $\o=3\O$ ($I^{THG}_x(\Omega)=I_0\left|  J^{NL}_{x}(3\Omega)\right|^2$) and we plot it as a function of temperature in Fig.\ 3e-f of the manuscript. As it is clear from Fig.\ \ref{fig-power}a-b, while for $\O=\O_2$ the temperature evolution of the THG intensity is similar for CP and Higgs processes, for $\O=\O_1$ the relative enhancement of the CP intensity at the temperature $T\simeq 0.9 T_c$, where the resonant condition $\O_2=\Delta_0(T)$ occurs, is much more pronounced with respect to the Higgs processes. When all the intensities are rescaled to the value obtained at the resonance, as done in Fig. 3d-f of the main text, this difference leads to the different $T$ evolution of the $I^{THG}(T)$ shown in the two panels e and f. 

\section{Derivation of the non-linear optical kernel within the pseudospin formalism}

Let us shown how the non-linear optical kernel $K_{ij}$ of Eq.\ \pref{kij} can be derived also by means of the Anderson pseudospin formalism, used in the previous theoretical work\cite{shimano14,TA}. The Anderson pseudospin are defined by $\mathbf{\sigma}_\mathbf{k}=\frac{1}{2}\Psi^\dagger\sigma\Psi_\mathbf{k}$, where $\Psi_\mathbf{k}=\left( c_{\mathbf{k}\uparrow}, c^\dagger_{-\mathbf{k}\downarrow} \right)^T$ is the Nambu spinor and $\sigma=\left(\sigma_1,\sigma_2,\sigma_3\right)$, so that the components of  $\mathbf{\sigma}_\mathbf{k}$ represent the amplitude, phase and density degrees of freedom, respectively.
The mean-field  Hamiltonian in the presence of the (uniform) electromagnetic field can then be written as:
\begin{equation}
	H_{BCS}=2\sum_\mathbf{k}\mathbf{b}_\mathbf{k}\cdot\mathbf{\sigma}_\mathbf{k}\quad,
\end{equation}
with:
\begin{equation}
\mathbf{b}_\mathbf{k}=\left( -\Delta'(t),-\Delta''(t),\frac{\xi_{\mathbf{k-e\mathbf{A}(t)}}+\xi_{\mathbf{k+e\mathbf{A}(t)}}}{2} \right)\quad,
\end{equation}
and $\Delta(t)=\frac{U}{N_s}\sum_\mathbf{k}\left\langle \sigma^x_\mathbf{k}(t)+i\sigma^y_\mathbf{k}(t)  \right\rangle$ the SC order parameter. Here the brackets $\langle\cdot\rangle$ denote the expectation value computed over the BCS ground state  $\left|\right.  \Psi_{BCS}  \left.  \right\rangle$, so that the previous definition of $\Delta$ is self-consistent.

The equation of motion for $\langle	\mathbf{\sigma}_\mathbf{k}(t)	\rangle$, as deduced from the Shroedinger equation for $\left|\right.  \Psi_{BCS}(t)  \left.  \right\rangle$, can be written as:
\begin{equation}\label{pseudospin_time_evolution}
 \partial_t\langle\mathbf{\sigma}_\mathbf{k}\rangle=2\mathbf{b}_\mathbf{k}\times\langle\mathbf{\sigma}_\mathbf{k}\rangle\quad.
\end{equation}

We solve \eqref{pseudospin_time_evolution} in perturbation theory with respect to $A^2$, by writing:
\begin{equation}
	\langle\mathbf{\sigma}_\mathbf{k}(t)\rangle=\langle\mathbf{\sigma}_\mathbf{k}(0)\rangle+\delta \mathbf{\sigma}_\mathbf{k}(t)\quad\text{and}\quad
	\mathbf{b}_\mathbf{k}(t)=\mathbf{b}_\mathbf{k}(0)+\delta\mathbf{b}_\mathbf{k}(t)\quad,
\end{equation}
with:
\begin{subequations}
	\begin{equation}
		\langle\mathbf{\sigma}_\mathbf{k}(0)\rangle=\left(\frac{\Delta_0}{2E_\mathbf{k}},0,-\frac{\xi_\mathbf{k}}{2E_\mathbf{k}}\right)\quad;
	\end{equation}
	\begin{equation}
		\mathbf{b}_\mathbf{k}(0)=\left(-\Delta_0,0,\xi_\mathbf{k}\right)\quad;
	\end{equation}
	\begin{equation}
		\delta\mathbf{b}_\mathbf{k}(t)=\left(-\delta\Delta'(t),0,\frac{e^2}{2}\sum_i\partial^2_i\epsilon_\mathbf{k}A_i^2(t)+O(A^4)\right)\quad,
	\end{equation}
	\end{subequations}
where we included for the sake of simplicity only the amplitude fluctuations $\delta\Delta'(t)$, the extension to the case having also phase fluctuations $\delta\Delta"(t)$ being straightforward. 
	The linearized equations of motion at $T=0$ then read:
	\begin{equation}\label{linear_pseudospin_time_evolution}
		\left\{
		\begin{matrix}
			\partial_t\delta\sigma^x_\mathbf{k}(t)=-2\xi_\mathbf{k}\delta\sigma^y_\mathbf{k}(t)\\\\
			\partial_t\delta\sigma^y_\mathbf{k}(t)= 2\xi_\mathbf{k}\delta\sigma^x_\mathbf{k}(t) +2\Delta_0\delta\sigma^x_\mathbf{k}(t)+ \\\\
			+\frac{1}{2E_\mathbf{k}}\left[ e^2\Delta_0\sum_i\partial^2_i\epsilon_\mathbf{k}A_i^2(t)-2\xi_\mathbf{k}\delta\Delta'(t)\right]\\\\
			\partial_t\delta\sigma^z_\mathbf{k}(t)=-2\Delta_0\delta\sigma^y_\mathbf{k}(t) 
		\end{matrix}
		\right.\quad.
	\end{equation} 
Notice that \eqref{linear_pseudospin_time_evolution} have to be solved self-consistenlty, by imposing: $\delta\Delta(t)=\frac{U}{N_s}\sum_\mathbf{k}\left[\delta\sigma^x_\mathbf{k}(t)+i\delta\sigma^y_\mathbf{k}(t)\right]$.

The first and the last equations of the system \eqref{linear_pseudospin_time_evolution}, along with the initial conditions $\delta\sigma_\mathbf{k}(0)=0$, lead to the identity: 
\be
\lb{iden}
\xi_\mathbf{k}\delta\sigma^z_\mathbf{k}(t)\equiv\Delta_0\delta\sigma^x_\mathbf{k}(t),
\ee
which allows one to reduce the number of equations. 

The non-linear current can be expressed in the pseudospin formalism as\cite{shimano14}:
\be
\lb{jnlta}
J^{NL}_i=-2e^2 A_i(t)\sum_\bk \frac{\pd^2\e_\bk}{\pd k_i^2} \delta \sigma_\bk^z(t), 
\ee
i.e. it is controlled by density fluctuations, in agreement with the general argument discussed in the main manuscript. To compute the current one then needs to deduce  $\delta\sigma_\bk^z(t)$ by solving the system \eqref{linear_pseudospin_time_evolution}. By simple algebra one can express it in the 
Fourier space as:
\bea
\delta \sigma_\bk^z(\omega)&=&e^2\sum_j \pd^2_j\e_\bk \frac{\Delta_0^2}{E_\bk(\o^2-4E_\bk^2)} A_j^2(\o)+\nn\\
\lb{deltaz}
&-&\frac{\Delta_0 \xi_\bk}{E_\bk(\o^2-4E_\bk^2)}\delta \Delta'(\omega),
\eea
where  $\delta \Delta'(\omega)$ from Eq.\  \eqref{linear_pseudospin_time_evolution}  coincides with  the expression \pref{deltaDELTA} derived above from Eq.\ \pref{SA}.
By inserting then Eq.\ \pref{deltaDELTA} into Eq.\ \pref{jnlta} one finds back the definition \pref{kij} of the electromagnetic kernel. In particular, one recovers again the presence of a CP term that is strongly diverging at $\omega=2\Delta_0$. This contribution has been overlooked by the authors of Refs.\ \cite{shimano14,TA} since they made the replacement
\be
\lb{ineq}
\sum_\bk \frac{\pd^2\e_\bk}{\pd k_i^2} \delta \sigma_\bk^z(t) \ra-\frac{1}{2} \sum_\bk \xi_\bk \delta \sigma_\bk^z(t).
\ee
By making the assumption \pref{ineq} and using the identity \pref{iden}, the authors of Ref.\ \cite{shimano14,TA} deduced that the non-linear current  should be always written as:
\begin{equation}
\lb{jpi4}
	J^{NL}(t)=\frac{e^2\Delta_0}{U}A(t) \langle\Delta(t)\rangle.
\end{equation}
However, the replacement \pref{ineq} is clearly {\em wrong}, since from Eq.s\ \pref{jnlta} and \pref{deltaz} one immediately sees that $J_i^{NL}$, being a {\em non-linear} current,  depends in general in asymmetric way on the two $A_i$ components. As we discuss in the main text, within the formalism of Eq.\ \pref{SA},  the replacement \pref{ineq} corresponds to replace each $\pd^2_i\e_\bk$ term in Eq.\ \pref{chi_0}  with $\xi_\bk$, thus removing the divergence of the CP part at $\omega=2\Delta$ and loosing the tensorial nature of the CP term. This mistake lead the authors of Ref.\ \cite{shimano14} to miss the existence of CP processes, and to attribute the THG intensity always to the sub-dominant Higgs-mode excitation.

\end{document}